\documentclass[preprint, aps]{revtex4-2}

\usepackage{svmacro}
\usepackage{comment}
\usepackage[T5]{fontenc}
\usepackage[utf8]{inputenc}
\usepackage{xcolor}

\def\la{\langle}
\def\ra{\rangle}

\def\be{\begin{equation}}
\def\ee{\end{equation}}
\DeclareMathOperator*{\sumsum}{\sum\sum}

\begin{document}

\title{Measurement time of weak measurements on large entangled systems}

\author{Trường-Sơn P. Văn}
\email{truongson.vanp@gmail.com}
\affiliation{Faculty of Applied Mathematics,
	Fulbright University Vietnam, Ho Chi Minh City, Vietnam}

\author{Andrew N. Jordan}
\affiliation{Institute for Quantum Studies, Chapman University, Orange, CA 92866, USA}
\affiliation{The Kennedy Chair in Physics, Chapman University, Orange, CA 92866, USA}
\affiliation{Department of Physics and Astronomy, University of Rochester, Rochester, NY 14627, USA}

\author{David W. Snoke}
\affiliation{Department of Physics and Astronomy, University of Pittsburgh, Pittsburgh, PA, 15260, USA}
\affiliation{Pittsburgh Quantum Institute,  Pittsburgh, PA, 15260, USA}

\begin{abstract}
	It is well established that starting only with strong, projective quantum measurements, experiments can be designed to allow weak measurements, which lead to random walk between the possible final measurement outcomes.  However, one can ask the reverse question: starting with only weak measurements, can all the results of standard strong measurements be recovered? Prior work has shown that some results can be, such as the Born rule for the probability of measurement outcomes as a function of wave intensity. In this paper we show that another crucial result can be reproduced by purely weak measurements, namely the collapse of a many-body, nonlocally entangled wave function on a time scale comparable to the characteristic time of a single, local measurement. For an entangled state of a single excitation among $N$ qubits, we find the collapse time scales as a double logarithm of $N$. This result affirms the self-consistency of the hypothesis that spontaneous weak measurements lie at the base of all physical measurements, independent of human observers.
\end{abstract}

\maketitle

\section{Introduction}
The measurement problem in quantum mechanics is one of the longstanding great questions of physics. In the standard formulation of quantum mechanics \cite{cohen-tann}, a ``strong'' measurement is defined as 1) the non-unitary projection of a quantum mechanical state onto one of the eigenstates of the operator that represents the measurement (known as a ``collapse of the wave function''), and 2) the subsequent normalization of the projected state to unity. The probability of collapse into any eigenstate is proportional to the square of the projected amplitude of the original state onto that eigenstate; this is known as the Born rule. The ``measurement problem'' arises from two debated questions. First, how can a non-unitary projection arise in a quantum mechanical universe that is described only by mathematics  (specifically, the many-body Schr\"odinger equation) with unitary time evolution? Second, what causes a specific outcome to occur in any given measurement--what selects one state out for collapse in any particular case?

Much has been written, of course, on various interpretations of this process, but over the past two decades, there has been great theoretical progress on the topic of ``weak measurement,'' \cite{jordan}, which operates entirely within the standard framework described above, but allows new questions to be asked about the nature of measurement. A ``weak'' measurement can be defined as having the following processes: 1) a quantum system, which we can call the ``internal'' system, is allowed to interact with an external, separate ``pointer'' system, which is macroscopically observable, such that the change made on the state of the pointer is much less than the uncertainty of the degree of freedom of the pointer state being observed;  2) subsequently, a standard, strong measurement is made of the pointer state (but not of the internal system), which collapses the pointer to a definite state.  If the internal system is in a superposition of states that have different effects on the pointer, the interaction with the pointer will create an entangled state of the internal system and the pointer.  The result of this is that the strong measurement of the pointer, which follows the Born rule, will effectively introduce random noise back into the behavior of the internal system. This noise will then lead to a random walk of the internal quantum system, which can be studied using standard statistics for random walks. There are some surprising results of this---for example, there is a finite probability that a random walk will reverse and retrace it path back to its starting point, which means that a partial measurement can be completely reversed without any dissipation \cite{korotkov2006undoing,katz2008reversal,jordan2010uncollapsing,jordan2}.

{While the above approach uses only standard quantum mechanical theory with projective measurement collapse, in recent years there has been interest in spontaneous collapse theory, which posits additional, non-unitary terms added to the Schr\"odinger equation that can give the same effect as weak measurement theory, namely collapse to definite projective states in accordance with the Born probability rule, but driven by a physical stochastic noise process. A well-studied version of this is the Ghirardi-Rimini-Weber (GRW) model \cite{GRW,bassi1,bassi3},  which posits universal noise from an external source.
More recently, a related but different approach has been proposed \cite{spont,spont2,spont3}, which posits just a non-Hermitian but norm-conserving operator of the form
\begin{equation}
	{\cal O}_W = i\varepsilon (\langle \sigma_z\rangle -\sigma_z )
	\label{ow}
\end{equation}
added to the Hamiltonian, where $\langle \ldots\rangle$ represents the expectation value over the full many-body quantum state, $\sigma_z$ has the form of the Pauli $z$-spin operator, but acting on the two fermionic states $|N=0\rangle$ and $|N=1\rangle$ of any fermionic eigenstate, and $\varepsilon$ fluctuates as a martingale due to actual, local environmental fluctuations. It has been shown \cite{spont,spont2,spont3} that this approach reproduces the Born rule for projective collapse rates and does not allow superluminal signaling. However, there is not yet consensus on whether this approach can reproduce all of the experimental results of standard quantum mechanics. In particular, the time scale for collapse via this mechanism has not been clear. If the time scale for collapse grows uncontrollably as the system size increases, it would be a strong argument that the proposed spontaneous collapse mechanism does not work.

In this paper, we explore this question, and show that the time scale for collapse for a large entangled system remains of the same order as the time scale for collapse of a single, unentangled system. In doing this, we show that the results for this spontaneous collapse model are exactly the same as for standard weak measurement theory. This, in turn, allows us to say that many weak measurements done on a large entangled system also will give collapse of the whole system on the time scale for collapse of a single measurement. This is a useful result for weak measurement theory, even if one does not commit to the proposed spontaneous collapse model.

We can see that the proposed non-unitary operator (\ref{ow}) will give the same results as weak measurement theory through the following simple argument \cite{spont3}.} 
Imagine a system with states of the form
\begin{eqnarray}
	|\psi\rangle = |\phi\rangle |X \rangle ,
\end{eqnarray}
where $|X\rangle$ is the state of an external pointer with a continuous range of states (e.g., the position of a needle in a meter),  and $|\phi\rangle$ is the state of a two-level internal system (i.e., a ``qubit''). We can write $|\phi\rangle = \alpha|-1\rangle +\beta|1\rangle$, where $\alpha$ and $\beta$ are complex coefficients of the two states $|-1\rangle$ and $|1\rangle$, respectively. { Taking $|X\rangle$ to be a Gaussian centered at $x = 0$ with uncertainty $\sigma$, after an interaction $\hat{V} =g\sigma_z (-i\hbar \partial/\partial x)$ with the pointer for a short time $dt$,} the state is
\begin{eqnarray}
	|\psi'\rangle &=&   \int dx  |x\rangle   \left( \alpha e^{-(x-gdt)^2/2\sigma^2}  |-1\rangle + \beta\ e^{-(x+gdt)^2/2\sigma^2}  |1\rangle \right),
\end{eqnarray}
where $g$ is a small parameter. After a strong measurement of the position $x$ of the pointer, the integral is removed, leaving only one $|x\rangle$ state, which we label $|x_0\rangle$. For $df = xgdt/\sigma^2 \ll 1$, the state is then (including the normalization to unity)
\begin{eqnarray}
	|\psi''\rangle   &=& \frac{1}{\sqrt{|\alpha|^2(1+df)^2 +|\beta|^2(1-df)^2 }}  \left( \alpha  (1+df) |-1\rangle + \beta  (1-df) |1\rangle\right) |x_0\rangle.
\end{eqnarray}
Simple math then shows \cite{spont3} that the time evolution corresponding to this change is given by
\begin{eqnarray}
	d|\psi\rangle = |\psi''\rangle - |\psi\rangle &=& x_0gdt  (\langle \sigma_z\rangle -\sigma_z ) |\phi\rangle|x_0\rangle, \label{weakop}
\end{eqnarray}
where $\sigma_z$ is the Pauli spin operator acting on the two quantum states of the internal system, and $x_0$ can be either negative or positive. This has exactly the same form as (\ref{ow}) above, even though it has been derived entirely from standard projective collapse theory.

The above analysis shows that the proposed addition (\ref{ow}) to the Schr\"odinger Hamiltonian is physically possible, since it is derived entirely from standard quantum mechanics in a physically possible weak measurement scenario. As such, we know that an operator of the form (\ref{ow}) cannot give superluminal signaling, or any other predictions that would violate standard projective measurement theory, when $\varepsilon$ fluctuates as a martingale.

 \subsection{Statement of the Problem}

As discussed above, both the weak measurement approach and spontaneous collapse using the operator ${\cal O}_W$, when applied to a two-state system, lead to a random walk on the Bloch sphere representing the two internal states, which eventually converges on either the top or the bottom of the Bloch sphere.
(For a review of the Bloch sphere representation, see, e.g., Ref.~\cite{snokeQM}, Section~14.3). Prior work (see e.g. \cite{spont3,jordan2,jordan2006qubit}) has shown that in the $t\rightarrow \infty$ limit, the probability of the random walk ending at one state or the other depends on the initial state in the same way as the Born rule for strong measurements; this is a very general result that does not depend on the details of the system other than that the random walk be a martingale walk, with no bias in one direction or the other. In the language of weak measurement, weak measurements extract only partial information from the internal system, but many weak measurements eventually have the same effect as a strong measurement.

Suppose now that we have a system with many internal states, instead of just a two-state qubit.  We can in general write any system with multiple states as
\begin{eqnarray}
	|\psi_m\rangle &=& \alpha_1 | 1\rangle + \beta_1\left(\frac{\alpha_2}{\beta_1}|2\rangle + \frac{\alpha_2}{\beta_1}|3\rangle  + \frac{\alpha_4}{\beta_1}|4\rangle +\ldots  \right) \nonumber\\
	&\equiv& \alpha_1|1\rangle + \beta_1| {\rm not}~ 1\rangle,
\end{eqnarray}
where $|\alpha_1|^2 + |\beta_1|^2 = 1$, and $n$ runs from 1 to $N$, with $\sum |\alpha_n|^2 = 1$. The random walk acting on the two states $|1\rangle$ and $|{\rm not}~1\rangle$ will have outcome $|\psi_m'\rangle =|1\rangle$ with probability $|\alpha_1|^2$. If the opposite happens,  and the system converges to $|\psi_m'\rangle = |{\rm not}~1\rangle$, a similar random walk among the remaining states with $n \ne 1$ will lead to convergence on state $|2\rangle$ with total probability $|\alpha_2|^2$; if instead it converges to $|{\rm not}~2\rangle$, then the same process can continue on state $|3\rangle$, and so on.

While this analysis shows that the Born rule applies for systems with $N > 2$ states when there is weak measurement on every state, it does not say anything about the time scale for the collapse into one outcome, however. In a standard strong measurement, one assumes that the collapse into one of the $N$ states will occur on the time scale of a single measurement, which is typically equated with the relevant decoherence time \cite{zurek}. It is not immediately obvious that a random walk among $N$ possible outcomes due to weak measurements will occur on the same  time scale; one might imagine something like motional narrowing occurring. Although we know that many weak measurements give the same final outcome (the Born rule) as one strong measurement \cite{spont3}, it might be the case that the time scale for collapse of $N$ entangled states is much longer than for a single strong measurement.

To address this question, we consider the concrete scenario of the detection of a charged particle in a cloud chamber or Geiger counter. A plane wave is prepared that has total intensity corresponding to exactly one photon.  This plane wave is then sent into a gas of $N$ atoms, each with a two-level transition that can be excited by that photon with an equal, small probability. After exposure to the plane wave, the quantum many-body state of the gas is
\begin{eqnarray}
	|\psi_m \rangle = \alpha_1|1,0,0,0,\ldots\rangle + \alpha_2|0,1,0,0,\ldots\rangle + \alpha_3|0,0,1,0,\ldots\rangle + \ldots +  \alpha_N|0,0,0,\ldots 0,1\rangle, \nonumber\\
	\label{mass_super}
\end{eqnarray}
where each of the states in the superposition is a Fock state corresponding to a single atom in an excited state. In the strong measurement scenario, a measurement of the entire gas of atoms (which is a local, macroscopic avalanche in a cloud chamber or Geiger counter) will occur in one step, collapsing the entire many-body state nonlocally to just one excited atom on the time scale of a measurement, independent of how far the gas extends in space.

We now imagine instead that each of the atoms simultaneously undergoes continuous weak measurement of its internal state.
As discussed above, the state (\ref{mass_super}) can be written in terms of two states for any given atom, e.g.,
\begin{eqnarray}
	|\psi_m \rangle &=& \alpha_1|1\rangle |0,0,0\ldots\rangle
	+ \beta_1|0 \rangle \left(\frac{\alpha_2}{\beta_1} |1,0,0,\ldots\rangle + \frac{\alpha_3}{\beta_1} |0,1,0,\ldots\rangle + \ldots \right. \nonumber\\
	&& \left. \hspace{5cm} + \frac{\alpha_N}{\beta_1} |0,0,\ldots ,0,1\rangle \right),
	\label{mass_super2}
\end{eqnarray}

{ As discussed above, we know that the system will obey the Born rule for collapse to any of the outcomes. However, we would like to compute the time for this to happen, on average. If, for example, the cloud chamber contains $10^{23}$ atoms, and the time for collapse into just one atom being excited scales as $N$, then if the time for one atom to absorb a photon is 1 picosecond, the time for $N$ entangled atoms to collapse will be of order $10^{11}$ seconds, that is, more than 3000 years.
One could even imagine a scenario in which the many weak measurements on the $N$ atoms  prevent the system from ever converging to a single collapse in the limit of $N \rightarrow \infty$, similar to a motional narrowing effect, or the ``quantum Zeno effect'' \cite{zeno}.

}

\subsection{The Evolution Equation for Entangled Collapse}

To track the process of collapse, we will keep track of the vertical component of the Bloch sphere, namely $U_{3,n} = |\alpha_n|^2-|\beta_n|^2 = 2|\alpha_n|^2-1$, for each atom $n$. The other components of the Bloch vector are assumed to remain coherent such that the length of the Bloch vector remains always unity; this is an explicit assumption of the spontaneous collapse model of Refs.~\cite{spont,spont2,spont3}, and Appendix~\ref{app:derivation} shows that it is also the case for weak measurements.

In this notation, the direct action of the operator $O_W$ on a single atom with two available states is
\begin{eqnarray}
	d U^{\rm direct}_{3,n} = g\omega_{R,n} (1- U_{3,n}^2)dt,
\end{eqnarray}
where $\omega_{R,n}$ is a Rabi rotation frequency playing the role of the parameter $\varepsilon$ introduced above, which gives the fluctuations acting on atom $n$. 
As seen in (\ref{mass_super2}), when $|\beta_{n}|^2 = (1- U_{3,n})/2$ changes, then each value of $|\alpha_{n'}|^2 = (U_{3,n'}+1)/2$ for $n'\ne n$ must change proportionately to keep the total norm the same.
The effect of this on other atoms is then found by computing the difference
\begin{eqnarray}
	dU^{\rm entangle}_{3,n'}  &=&  (U_{3,n'}+1)\frac{1-(U_{3,n}+d U^{\rm direct}_{3,n})}{1-U_{3,n}}- ( U_{3,n'}+1) .
\end{eqnarray}
This can then be mathematically rewritten as
\begin{eqnarray}
	dU^{\rm entangle}_{3,n'}
	&=&  \frac{U_{3,n'}+1}{1-U_{3,n}} ({1-U_{3,n}-dU^{\rm direct}_{3,n}})- (U_{3,n'}+1) \nonumber \\
	&=& - dU^{\rm direct}_{3,n} \frac{1+U_{3,n'}}{1-U_{3,n}} \nonumber \\
	&=& -g\omega_{R,n}(1+ U_{3,n})(1+U_{3,{n'}})dt.
	\label{atomnp}
\end{eqnarray}
We can call this the ``entanglement'' term, since it gives a change of the wave function affecting each atom dependent on the states of the other atoms.

The sum of the direct and entanglement terms then gives
\begin{eqnarray}
	\frac{\partial U_{3,n}}{\partial t}
	= \omega_{R,n} (1- U_{3,n}^2)
	- \sum_{n' \ne n} \omega_{R,n'}(1+ U_{3,n})(1+U_{3,{n'}}).
	\label{atomn}
\end{eqnarray}
It is easy to show that the second, entanglement term ensures conservation of the norm of the wave function:
\begin{equation}
 \sum_n (\partial U_{3,n}/\partial t) = 2({\partial /\partial t}) {\displaystyle  \sum_n} |\alpha_{n}|^2 = 0. 
 \end{equation}

While the result (\ref{atomn}) has been derived using the operator $O_W$, it can also be derived from standard weak measurement theory, as shown in Appendix A. In what follows, since the mathematical formalism is the same, the results do not distinguish between the weak measurement interpretation and the physical spontaneous collapse interpretation.

\section{Simulating the random walk}
\label{sec:sim}

We are now in a position to numerically model the evolution of a system with $N$ entangled states, and find the average time to collapse.
The conservation of the norm of the full wavefunction plays a crucial role in giving an effective interaction between all the atoms.

We assume that the initial many-body state $(\alpha_{0,1}, \dots, \alpha_{0,N})$ has equal probability for all states, $|\alpha_{0,n}|^2 = 1/N$, which implies $U^{0}_{3,n} = 2\left| \alpha_{0,n} \right|^2  -1.$
We can model the time evolution of (\ref{atomn}) as discrete steps $i$, such that $U^i_{3,n}$ satisfies
\begin{equation}
	U^{i+1}_{3,n} = U^i_{3,n} + \frac{\partial U_{3,n}(t_i) }{\partial t} \Delta t,
	\label{eq:discrete}
\end{equation}
and replace $\omega_{R,n}(t_i)$ with $X_{n}^i/(\Delta t)^{1/2}$.
Here, $\left\{ X_n^i \right\}_{i,n \in \N}$ are independent identically distributed (I.I.D.) random variables, sampled from a symmetric
distribution on $\R$. We used three different symmetric distributions with standard deviation $1$, namely the
normal distribution $\mathcal N(0,1)$, the
discrete Bernoulli distribution $\Ber\left(\{-1,1\}, \frac12\right)$, and the uniform distribution $\mathcal U\left([-\sqrt3,\sqrt3]\right)$.
The factor $(\Delta t)^{1/2}$ 
is introduced so that the limit $\Delta t \rightarrow 0$ can be obtained while keeping the effective diffusion constant $D = l^2/\tau = 1$, where $l$ is the average of the step size $|\omega_{R,n}|\Delta t$,  and $\tau = \Delta t$ is the characteristic time of the steps. With this convention, the units of time in all of the simulations below are $1/D$, which is the characteristic time for collapse in the case of $N=2$, i.e., a single unentangled qubit.

It is a standard result in It\^o calculus (see Ref.~\cite{Higham2021, Karatzas1991}) that under these assumptions,
Equation~\eqref{eq:discrete} converges to the stochastic differential equation
\begin{equation}
	dU_{3,n} = \left( 1 - U_{3,n}^2  \right) dW^n_t
	- \sum_{n'\not= n}^N (1 + U_{3,n}) (1 + U_{3,n'})  dW^{n'}_t \,,
	\label{eq:sde}
\end{equation}
where $W= (W^1, \dots, W^N)$ is the $N$-dimensional Wiener process.
{
To the best of our knowledge, the stochastic differential equation~\eqref{eq:sde} representing weak measurement of entangled states is new, and its exit time has not been analyzed before in the literature.
This equation can also be generalized by a more general weak measurement theory (see Appendix~\ref{app:derivation}).
}

\begin{figure}[b]
	\centering
	\includegraphics[width =0.8\linewidth]{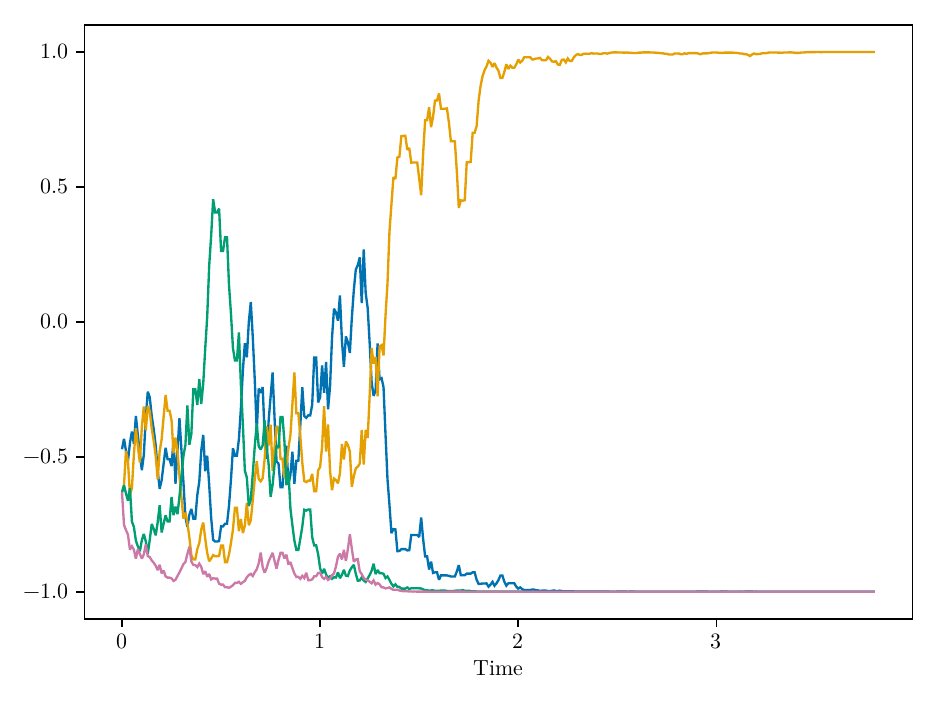}
	\caption{An example of the evolution for $N = 4$, $\Delta t = 1/100$. Each trajectory is a component $U_{3,n}$ where $n= 1,2,3,4$.
		One component is attracted to $1$ while the rest collapse to $-1$.}
	\label{fig-path}
\end{figure}

We used the Euler-Maruyama method to perform numerical simulations via Julia language. It must be noted that $U_{3,n}$ will never strictly hit either $-1$ or $1$ in finite time.
However, in our simulation, for a fixed $N$,
the system will stabilize to just one state having $\alpha_n \simeq 1$ in finite time (see Figure~\ref{fig-path}).
Although the system will not hit exactly $|\alpha_n| = 1$, the form of (\ref{atomn}) implies that there are basins of attraction around $-1$ and $1$, so
that once any of the $U_{3,n}$'s gets trapped
in either of them, it will stay there with high probability. We therefore define a ``collapse'' as when any one of the atoms reaches $U_{3,n} = 1 - \delta$, where $\delta \ll 1$ (see Appendix~\ref{app:stability} for a proof of this phenomenon).
Our simulation shows that, as the number of entangled atoms $N$ grows to infinity,
the collapse time $T$ grows very slowly, as $T \sim \ln \ln N$ (see Figure~\ref{fig-time}).

A heuristic argument for this phenomenon is given in Appendix~\ref{app:heuristics}.
Because of the vanishing correlations among the $U_{3,n}$'s (Theorem~\ref{theorem1}),
we can assume that they are independent.
Furthermore, we can also assume that initial fluctuations of all trajectories
are of similar size and that $U_{3,n} \sim 1/T - 1$ after some short initial time period (see~\eqref{eq:initial}).
As a consequence, by using Taylor expansion of the solution of the equation~\eqref{eq:sde},
we are able to relate the the random variables $U_{3,n}$ to the law of iterated log,
which gives us the $\ln \ln N$ behavior.

\begin{figure}[t]
	\centering
	\includegraphics[width =0.8\linewidth]{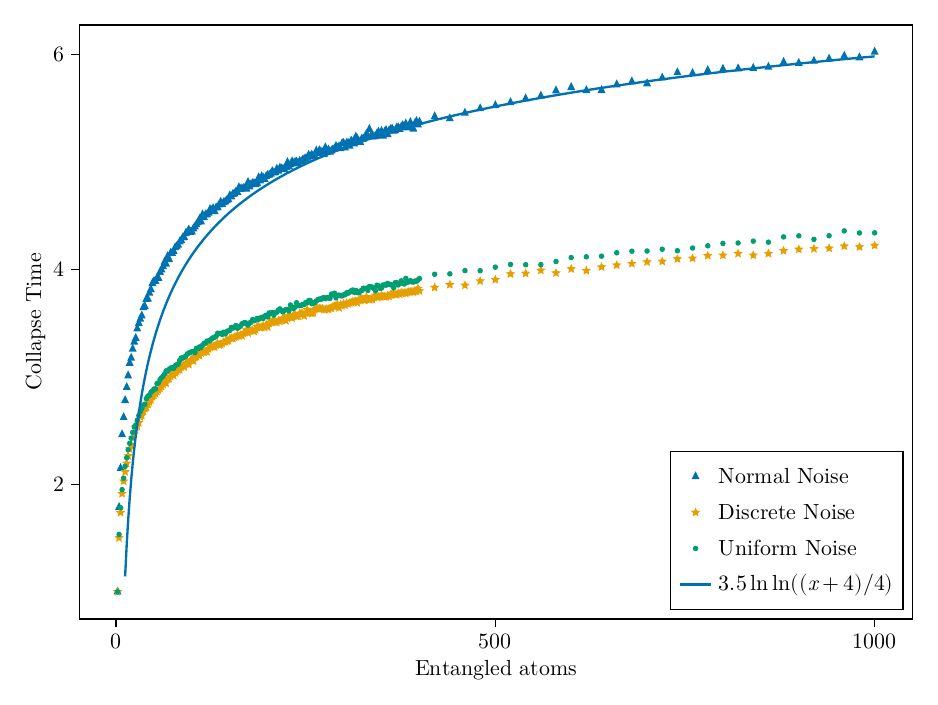}
	\caption{Collapse time for growing $N$ from 1 to 1000, where the case of $N=2$ (a two-state collapse, i.e., a single qubit) corresponds to collapse time of 1, for $\delta = 10^{-2}$.
		In all of the plots, $\Delta t = 1/25$ and the number of realizations to average over is 10000.}
	\label{fig-time}
\end{figure}

\section{Dynamics from Weak Measurement}
In our considered model, there is no intrinsic Hamiltonian dynamics of the system, 
so we can directly use the theory of continuous quantum measurement (see Chapter 5 in Ref.~\cite{jordan}) and the quantum Bayes rule to find the exact solution to the dynamics of the entangled state.
Recall the single-excitation entangled state (\ref{mass_super}). 
Let us perform a weak measurement of the vertical component of the Bloch sphere
representation of the state.
The measurement readout is \emph{on average} defined to be $1$ for excited state and $-1$ for non-excited state, in appropriate units.  
However, there is also detector noise added to the signal, which masks the true value of the quantum observable, 
making the measurement weak and continuous. 
Thus, the measurement readout can in principle take any value on the real line, although it will be concentrated around $1$ or $-1$. 
We assume Gaussian measurement statistics for the readout,
i.e., the distribution for an outcome $r$ (a continuous random variable)  for duration $dt$, conditioned on a postselection state $s$ ($s = \pm 1$)
is given by the Gaussian distribution
\begin{equation}
	P(r) = \left( \frac{dt}{2\pi \tau_m}  \right)^{1/2}  \exp\left( -\frac{dt}{2 \tau_m} (r - s)^2  \right)\,,
\end{equation}
where $\tau_m$ is the characteristic measurement time. 

Then, for $N>1$ independent measurements on the qubits in each of $N$ atoms, 
the collective distribution of outcomes $r_1, r_2, \ldots r_N$, for duration $dt$ provided 
the system is prepared in a single state with one excitation in site $n$ then is given by
\be
P_n(r_1, \ldots r_N) = \left(\frac{dt}{2\pi \tau_m}\right)^{N/2}
\exp\left(-\frac{dt}{2\tau_m} \left( (r_n-1)^2 + \sum_{j \ne n} (r_j+1)^2  \right) \right).
\ee
We repeat these measurements successively over time.

By the law of total probability, the total distribution of all measurement results is given by
\be
P_{tot}(r_1, \ldots r_N) = \sum_{n=1}^N |\alpha_n|^2 P_n(r_1, \ldots, r_N).
\ee
Given measurement results $r_1, r_2, \ldots, r_N$, we can write the conditional state of the quantum system via the Quantum Bayes rule \cite{korotkov1999continuous,jordan} as
\be
\label{eq:weakdynamic}
\alpha_n(t+dt) = \frac{\sqrt{P_n(r_1, \ldots, r_N)}}{\sqrt{P_{tot}(r_1, \ldots, r_N)}} \alpha_n(t).
\ee
Here we assume the simplest type of measurement where additional phases are not added by the measurement apparatus; a so-called quantum nondemolition (QND) measurement.
After canceling the quadratic terms in $r$, we define an unnormalized coefficient ${\tilde \alpha}$.  Expanding to first order in time, we find the equation of motion
\be
\frac{d}{dt}{\tilde \alpha}_n =\frac{1}{2 \tau_m} \left( -\sum_{j \ne n} r_j + r_n\right) {\tilde \alpha}_n,
\ee
which has the formal solution
\be
{\tilde \alpha}_n(t) ={\tilde \alpha}_n(0) \exp\left(  \frac{1}{2 \tau_m} \int_0^t dt' \left( -\sum_{j \ne n} r_j(t') + r_n(t') \right) \right) = {\tilde \alpha}_n(0) \exp\left(  \frac{1}{2} \left( -\sum_{j \ne n} R_j + R_n \right) \right).
\ee
Here we see that because of the QND nature of the measurement, only the time-integrated measurement signals matter for the quantum state dynamics.  We define a new variable 
\begin{equation}
	R_n  = \frac{1}{\tau_m}\int_0^t dt' r_n(t') 
\end{equation}
to simplify notation.  Therefore the conditional (normalized) quantum state dynamics is given by $\alpha_n(t) = {\tilde \alpha}_n(t) /\sqrt{ \displaystyle \sum_n |{\tilde \alpha}_n(t)|^2}$. In the spontaneous collapse interpretation, one can simply treat the variables $R_j$ as fluctuations of the environment without reference to auxiliary measuring degrees of freedom.  We note that we can further exploit the unnormalized states to write a simpler form by multiplying all the state coefficients by a common factor of $\exp((1/2){\displaystyle \sum_j }R_j)$, to get a simple collapse equation
\be
{\alpha'_n}(t) = { \alpha'}(0) \exp(R_n).
\ee
As one of the $R$ values increases in time, the others decrease, so there is an exponential separation of the coefficients, causing collapse of the normalized state.

The total probability density of the time-integrated results $R_1, R_2, \ldots R_N$ interpreted as random variables is given by
\be
P(R_1, \ldots R_N) = \sum_n |\alpha_n(0)|^2  \left(\frac{\tau_m}{2\pi t }\right)^{N/2}
\exp\left(-\frac{\tau_m}{2 t} \left(\sum_{j \ne n} (R_j+t/\tau_m)^2 + (R_n-t/\tau_m)^2 \right) \right).
\ee
That is, the average of outcome $R_j$ is given by $\la R_j \ra = |\alpha_j|^2 t /\tau_m - {\displaystyle \sum_{k\ne j}} |\alpha_k|^2 t/\tau_m$, with variance of any outcome given by $t/ \tau_m$; the signal will grow linearly in time if the excitation is on the right detector; otherwise it will decrease linearly in time.  This signal is obscured by noise whose standard deviation grows as square-root of time.  This gives the complete solution to the problem from which one can investigate other quantities of interest.

 We define the criterion for collapse to a given excitation to be $2 |\alpha_n(t)|^2 - 1 = 1 -\delta$, where $\delta$ is very small.  Expressing this condition in terms of the unnormalized states, we require that for some index $n$, we have the condition
\be
|{\tilde \alpha}_n(t)|^2 = \frac{1-\delta/2}{\delta/2}  \sum_{j \ne n}
|{\tilde \alpha}_j(t)|^2.
\ee
Inserting our exact results converts this condition to
\be
\label{eq:conditionR}
-2 R_n\ge \ln \frac{1-\delta/2}{\delta/2} + \ln \sum_{k \ne n} \exp\left(- 2 R_k \right).
\ee
As shown in Appendix~\ref{app:derivation}, the dynamics given by Equation~\eqref{eq:weakdynamic} is equivalent to
the dynamic given by~\eqref{eq:sde}.
Therefore, Theorem~\ref{theorem1} in Appendix~\ref{app:stability} applies and we deduce that it must be the case
over some period of time, that all but one of the $\alpha_n$'s will vanish and condition~\eqref{eq:conditionR} will be met.

\section{Discussion}
We have seen that the time for collapse of the whole system into just one of the atoms being excited occurs in a time of the order of the collapse time for a single two-level system, i.e., the decoherence time, but with a very slow, $\ln\ln N$ increase.
The implication of this is that even for a very large system, of the order of $N= 10^{23}$, $\ln\ln N \sim 4$; for all the baryons in the visible universe, $\ln \ln 10^{80} \sim 5$. Therefore, many weak measurements reproduce the effect of a single strong measurement, not only in the probability statistics of the Born rule, but also with a time scale for measurement of the order of the decoherence time of a single atom.

In weak measurement theory, the weak measurements are typically assumed to require detection with an observer; that is, weak measurements are taken as a particular limit proven to be physically possible within the standard Copenhagen model of strong measurement.  The results discussed here, however, put us in a position to ask whether all strong measurements could actually be the result of spontaneous collapse that occur via physical fluctuations according to the model of Refs.~\cite{spont,spont2,spont3}, even when there is no observer.  There are three ways we can imagine this could occur.  One way would be if there are continuous, universal fluctuations that give an operator of the form ${\cal O}_W$ with a time-varying $\varepsilon(t)$ at every point in space, as has been proposed, e.g., by Penrose, Di\'osi and coworkers \cite{penrose,diosi} on the basis of gravity, and generically in the scenario of continuous spatial localization \cite{csl}. This would give decoherence and collapse in the vacuum of free space.  This appears to be disfavored by recent experiments \cite{bassi,bassi2}.

A second possibility is that the $\varepsilon(t)$ at every point in space is due to fluctuations of the local environment due to actual physical inhomogeneity in the distribution of particles in the local environment. This has the advantage that it only gives collapse of the wave function due to decoherence, which meshes well with the work of Zurek and coworkers \cite{zurek} on ``einselection.'' If the fluctuations in the local environment have generic martingale form (i.e., are not biased toward one outcome), then the outcome would be the same as in the universal fluctuation scenario, but with a time scale that depended on the local fluctuations, and with no collapse in free space.

This scenario raises the possibility, however, of using some physical factor to bias the local fluctuations, so that they would not have martingale form. In this case, at least in principle, the Born rule could be violated, and violations of the Born rule could be used for superluminal communication. Deterministic spontaneous collapse has been shown to allow superluminal communication, while stochastic martingale fluctuations do not (see, e.g., Ref.~\cite{gisin} and Appendix A of Ref.~\cite{spont3}.)

A third possibility is that $\varepsilon(t)$ is proportional to the product of both of the above two types of fluctuations. In this case, superluminal communication would not be possible, but the time for collapse would still depend on the rate of local environmental fluctuations, and collapse would not occur for single particles in the vacuum of free space. In this case, there would not be a physical test to distinguish between continuous weak measurements and the standard measurement hypothesis. The appeal of this scenario would be primarily in allowing a self-consistent solution of the measurement problem without the need to define consciousness, knowledge, and so on.

The above analysis indicates that every strong measurement could be accounted for as the end result of many weak measurements. While it does not require a spontaneous collapse hypothesis, it shows that a physical spontaneous collapse mechanism of the form (\ref{ow}) does not give unduly long collapse times. While we do not know if any of the above three mechanisms exists for the fluctuation of $\epsilon(t)$, we can say that this spontaneous collapse mechanism is physically possible, because the above analysis uses only the formalism of weak measurement, which in turn is based on standard measurement theory.

	{\bf Acknowledgements}. We thank D.~Maienshein and G.~Iyer for helpful conversations, and A.~Gadepalli for help with the initial coding.
ANJ acknowledges support from the John Templeton Foundation Grant ID 63209.
This work has been supported in part by the Pittsburgh Quantum Institute.

\newpage
\appendix

\section{Derivation of the weak measurement evolution of entangled states}
\label{app:derivation}

The result (\ref{atomn}) can be seen as the particular case of a two-level system in the context of more general weak measurement theory, as follows.
We consider the most general form of the stochastic master equation of a normalized density matrix in It\^o form (Ref.~\cite{jordan}, Eq.~(9.79)),
\begin{eqnarray}
	d{\hat \rho} &=& - i [{\hat H}, {\hat \rho}]dt + \sum_{\nu=1}^m \left( {\hat L}_\nu {\hat \rho} {\hat L}^\dagger_\nu dt
	- \frac{1}{2} ({\hat L}^\dagger_\nu {\hat L}_\nu {\hat \rho} + {\hat \rho} {\hat L}^\dagger_\nu {\hat L}_\nu)dt \right.\\
	&+& \left. \sqrt{\eta_\nu} \left[ ({\hat L}_\nu {\hat \rho} + {\hat \rho} {\hat L}^\dagger _\nu) - {\rm Tr} [ {\hat L}_\nu {\hat \rho} + {\hat \rho} {\hat L}^\dagger _\nu] {\hat \rho} \right] dW_\nu\right). \nonumber
\end{eqnarray}
The corresponding shifted and rescaled readouts are given by
\begin{equation}
	r_\nu = \sqrt{\eta_\nu} {\rm Tr}[ ( {\hat L}_\nu +  L^\dagger_\nu){\hat \rho}] + \frac{dW_\nu}{dt},
\end{equation}
where the index $\nu = 1, \ldots, m$.
Here we have the most general Markovian case of $m$ readouts with Lindblad operators ${\hat L}_\nu$, Wiener increment\index{Wiener increment}s $dW_\nu$ and detection efficiencies $\eta_\nu$.

We begin with the case where there are $N$ qubits prepared in state with density marix $\hat\rho$, with a direct sum Hamiltonian ${\displaystyle {\hat H} = \sum_{j=1}^N }{\hat H}_j$ that is non-interacting.  We also consider $N$ independent detectors monitoring each of the $N$ different qubits.  We let each detector monitor the $\sigma_z$ Pauli operator of each qubit.  In this case, we specialize to the idealized case of perfect detection efficiency ($\eta_\nu =1$) and the Lindblad operators take the form
\be
{\hat L}_j = \frac{\sigma_z^{(j)}}{2 \sqrt{\tau_m}},
\ee
where we combine the detector and qubit labels into one label $j$.  Here $\tau_m$ is the characteristic measurement time of each detector, which we take to be identical for simplicity \cite{jordan}.  In this case, the measurement readout for each qubit takes for form
\begin{equation}
	r_j = \frac{z_j}{\sqrt{\tau_m}} +  \frac{dW_j}{dt},
\end{equation}
indicating that the detectors are continuously monitoring the $z$ qubit coordinate for each qubit.  We note that the readout can be rescaled to be ${\tilde r}_j = z_j + \sqrt{\tau_m} dW_j/dt$ in order to be directly proportional to the qubit coordinate, plus detector noise.

We can now monitor the Bloch coordinate for each qubit $(x_j, y_j, z_j)$, where $x = {\rm Tr}[\hat\rho\sigma_x]$ , $y ={\rm Tr}[\hat\rho\sigma_y]$, and $z= {\rm Tr}[\hat\rho\sigma_z]$ for the three Pauli matrices.  We let each qubit Hamiltonian be of the form
\be
H_j = \frac{E}{2} \sigma_z^{(j)} + \frac{\Delta}{2} \sigma_x^{(j)}.
\ee
Making these simplifications and taking the trace of the equation of motion with respect to $(\sigma_x^{(j)}, \sigma_y^{(j)}, \sigma_z^{(j)})$, we find for the $z$ equations of motion
\be
dz_j = \Delta y_j dt
+\frac{1}{\sqrt{\tau_m}}(1-z_j^2) dW_j
+ \sum_{i \ne j}^N \frac{1}{\sqrt{\tau_m}}
\left[ \la \sigma_z^{(i)} \sigma_z^{(j)} \ra - z_i z_j\right]dW_i. \label{zj}
\ee
Here we see there arises a contribution to the diffusion term that involves the cross-correlation between the $\sigma_z$ operators.  We define the correlation function as
\be
\la \sigma_z^{(i)} \sigma_z^{(j)} \ra = {\rm Tr}[\rho \sigma_z^{(i)} \sigma_z^{(j)}]. \label{corr}
\ee

In the main text, attention is restricted to single-excitation-type states, with no tunneling term ($\Delta = 0$).  For pure states with only one excitation, $|\Psi\ra = {\displaystyle \sum_{k=1}^N} c_k |k\ra$, where $|k\ra$ has one excitation only, the expectation value
\be
\la \sigma_z^{(i)} \sigma_z^{(j)} \ra = \la \Psi | \sigma_z^{(i)} \sigma_z^{(j)} |\Psi\ra,
\ee
is given by
\be
{\displaystyle \sum_{k\ne i,j}} |c_k|^2 - |c_i|^2 - |c_j|^2  = 1 - 2(|c_i|^2 + |c_j|^2) = -z_i -z_j -1,
\ee
where we used the fact that $\sigma_z$ flips the sign of the state with an excitation as well as the normalization of the total state.  This result indicates that the relevant spins are anti-correlated.  In this case, the solution (\ref{zj}) simplifies to
\be
dz_j = \Delta y_j dt
+\frac{1}{\sqrt{\tau_m}}(1-z_j^2) dW_j
- \sum_{i \ne j}^N
(1+z_i)(1+z_j)\frac{dW_i}{\sqrt{\tau_m}}, \label{zj2}
\ee
which is equivalent to the result in the main text.

This calculation also shows that the conditional state of the quantum system remains coherent during the continuous collapse process.  To that end, we can also find the equations of motion for the coherence of the Bloch coordinates.  Taking the trace of the equation of motion with each of the other Pauli matricies, we find
\be
dx_j = -E y_j dt - \frac{x_j}{2 \tau_m} dt - \frac{x_j z_j}{\sqrt{\tau_m}} dW_j + \sum_{i \ne j} \left( \left\langle \frac{\{ \sigma_x^{(j)}, \sigma_z^{(i)}\}}{2}\right\rangle - x_j z_i \right) \frac{dW_i}{\sqrt{\tau_m}}.
\ee
Here we have used the anti-commutator notation for the $x-z$ correlation function, defined similarly to the $z-z$ correlation function (\ref{corr}).

When applied to the special case of the single excitation subspace, the expectation of the anti-commutator vanishes because the action of the $\sigma_x$ operator results in either no excitation, or two excitations, which has no overlap with the original subspace.

Similarly, for the $y_j$ dynamics we find the result
\be
dy_j = E x_j dt - \Delta z_j dt - \frac{y_j}{2 \tau_m} dt - \frac{y_j z_j}{\sqrt{\tau_m}} dW_j + \sum_{i \ne j} \left( \left\langle \frac{\{ \sigma_y^{(j)}, \sigma_z^{(i)}\}}{2}\right\rangle - y_j z_i \right) \frac{dW_i}{\sqrt{\tau_m}}.
\ee
Just as with the $\sigma_x$ operation, when applied to the single excitation subspace, the correlation function vanishes.

It is of interest to investigate how the continuous measurement process purifies each qubit state.  Initially, the entangled nature of the total state makes each reduced qubit into a mixed state ($x_j^2+y_j^2+z_j^2 <1)$.  As the measurement progresses, the continuous collapse process eventually projects one of the qubits into an excited state (1, or $z=-1$), while the rest are projected into the ground state (0, or $z=+1$).  Any such state is separable and pure.  Therefore, the purity of the reduced state must {\it increase} during the measurement process.  We can calculate the average increase of purity for the jth qubit via its definition $P_j = x_j^2+y_j^2+z_j^2$.  We calculate the stochastic change in the purity via $dP = 2 x dx + 2 y dy + 2 z dz + dx^2 + dy^2 + dz^2$, where we go to second order and apply It\^o's rule $dW_i dW_j = dt \delta_{ij}$.  Taking the stochastic average drops the remaining stochastic terms leaving the result
\be
\la dP_j \ra =  \left[ (1-P_j)(1-z_j^2) +
	(1+z_j)^2 \sum_{i \ne j} (1+z_i)^2 + (P_j - z_j^2) \sum_{i \ne j} z_i^2 \right] \frac{dt}{\tau_m}.
\ee
The first contribution in the average purification has been well studied in the past; see e.g. Refs.~\cite{jacobs2003project,combes2006rapid,jordan2006qubit}.  We see the fastest rate of purification is at $z_j=0$.  The other contribution comes from the information gain from the other measurements.  The fact that they are positive leads to an increased projection rate of the reduced qubit.

\section{Proof that $N$-state system stabilizes}
\label{app:stability}

\begin{theorem}
	\label{theorem1}
	Consider Equation~\eqref{eq:sde} for any $N > 2$.
	For any pair $1\leq n,k\leq N$, $n\not= k$, it is true that
	\begin{equation}
		\lim_{t\to \infty } \E (U_{3,n}+1) (U_{3,k} + 1)  = 0 \, .
	\end{equation}
\end{theorem}
Here $\E$ indicates the expectation value.
This theorem implies that, as $t\to\infty$, with high probability, $(U_{3,n}+1) (U_{3,k} + 1) \to 0$ ``somewhat monotonically,''
meaning, with high probability, one of the $U_{3,n}$'s will be trapped arbitrarily near 1 while all the other $U_{3,k}$'s near $-1$.

\begin{proof}
	Consider Equation~\eqref{eq:sde}.
	For convenience, letting  $V_n = U_{3,n} + 1 \in [0,2]$, we have the system
	\begin{equation}
		\begin{dcases}
			dV_n = V_n (2 - V_n) \, dW^n_t - \sum_{k\not= n}^N V_n V_k dW^k_t \,, \\
			\sum_{n=1}^N V_n = 2 \,.
		\end{dcases}
		\label{eq:reformulate}
	\end{equation}
	Therefore, the quadratic variation of $V_n$ satisfies
	\begin{align}
		\frac{d}{dt} \langle V_n \rangle & =  V_n^2 (2- V_n)^2 + \sum_{k\not=n}^N V_n^2 V_k^2   \nonumber \\
		                                 & \geq V_n^2 (2 - V_n)^2  \,.
	\end{align}

	By the It\^o formula, we have
	\begin{equation}
		d V_n^2 = 2 V_n dV_n + \langle V_n \rangle dt \,.
	\end{equation}
	Therefore,
	\begin{equation}
		\frac{d}{dt} \E V_n^2 =  \E \langle V_n \rangle \geq
		\E \left(V_n^2 (2 - V_n)^2 \right)            \,,
		\label{ine:expected}
	\end{equation}
	which implies
	\begin{align}
		\frac{d}{dt} \E \left( \sum_{n = 1}^N V_n^2 \right)
		 & \geq    \sum_{n = 1}^N \E \left( V_n^2 (2 - V_n)^2 \right)                 \nonumber      \\
		 & =  \sum_{n = 1}^N \E \left( V_n^2 \left(\sum_{k\not=n}^N V_k\right)^2  \right)  \nonumber \\
		 & \geq     \sumsum_{\substack{n,k=1                                                         \\n\not=k}}^N\E \left( V_n^2 V_k^2 \right)
		\geq \frac{1}{N^2} \E \left(\sumsum_{\substack{n,k=1
		\\n\not=k}}^N V_n V_k \right)^2  \,.
		\label{ine:lower}
	\end{align}
	Because $\displaystyle \sum_{n=1}^N V_n = 2$, it is true that
	$\displaystyle \sum_{n=1}^N V_n^2  = 4 - \sum_{\substack{ n,k =1 \\ n\not= k}}^N V_k V_n$.
	Substituting this and into~\eqref{ine:lower}, we get
	\begin{equation}
		\frac{d}{dt} \E \left(\sumsum_{\substack{ n,k =1 \\ n\not= k}}^N V_k V_n  \right)
		\leq -
		\frac{1}{N^2} \E \left(\sumsum_{\substack{n,k=1 \\n\not=k}}^N V_n V_k \right)^2
		\leq -
		\frac{1}{N^2} \left(\E \sumsum_{\substack{n,k=1 \\n\not=k}}^N V_n V_k \right)^2\,.
	\end{equation}
	Thus,
	\begin{equation}
		\sumsum_{\substack{ n,k =1 \\ n\not= k}}^N \E (V_k V_n)  \leq \frac{1}{(t/N^2 + C)} \,.
		\label{ine:correlation}
	\end{equation}
	Because $V_n \geq 0$ for all $1\leq n \leq N$,
	this implies the statement of Theorem~\ref{theorem1}.
\end{proof}

\begin{remark}
	If $V^n_0 = 2/N$ for all $n= 1, \dots, N$,
	by symmetry, we may expect that $V^n$'s are exchagneable. Therefore,
	we may pick $C = 1/4$ for the inequality to hold at time $0$. The inequality then becomes
	\begin{equation}
		\E (V_k V_n) \leq \frac{4}{ 4t + (N-1)^2}\,.
		\label{ine:remark}
	\end{equation}
\end{remark}

This result is stronger than the result in Appendix B of~\cite{spont3},
which only shows that the process gets near the end points $\set{-1,1}$, but does not address their stability. The stability result here justifies the cut-off of size $\delta$ near $1$ used in the our numerical simulations.

\section{Heuristics for $\ln \ln N$ bound}
\label{app:heuristics}

Considering~\eqref{eq:reformulate}, by It\^{o} formula, we have
\begin{align}
	d \ln (V^n_t) 
	&=
	\frac{d V^n_t}{V^n_t} - \frac{1}{2(V^n_t)^2}\left((V^n_t)^2 (2 - V^n_t)^2 
		+ \sum_{k\not= n} (V^n_t)^2 (V^k_t)^2 \right) \, dt \nonumber \\
	&=  (2 - V^n_t) \, dW^n_t + \sum_{k\not=n}^N  V^k_t \, dW^k_t 
	- \frac{1}{2}\left( (2 - V^n_t)^2 
		+ \sum_{k\not= n}^N (V^k_t)^2 \right) \, dt  \,.
\end{align}

Therefore, letting 
\begin{equation}
Q_t = \frac{1}{2}\left( (2 - V^n_t)^2 
+ \sum_{k\not= n}^N  (V^k_t)^2 \right) \,,
\end{equation}
we have
\begin{equation}
	V^n_t  =
	V^n_0 \exp\bigg(\int_0^t 2W^n_t   
	- \sum_{k=1}^N \int_0^t V^k_s \, dW^k_s - \int_0^t Q_s \, ds  \bigg) \,.
	\label{eq:solutionV}
\end{equation}
Assuming $V_0^n = 2/N$, expanding the above to the second order Taylor expansion and letting $X_k = \int_0^t V^k_s \, dW^k_s$, we have
\begin{align}
	V^n_t & \sim
	V^n_0 + V^n_0 \left( W_t^n - \sum_{k=1}^N X_k - \int_0^t Q_s \, ds  \right)
	+ \frac{V^n_0}{2} \left( W_t^n - \sum_{k=1}^N X_k  - \int_0^t Q_s \, ds\right)^2      \nonumber \\
	      & =
	\frac{2}{N} + \frac{2}{N} \left( W_t^n - \sum_{k=1}^N X_k  - \int_0^t Q_s \, ds\right)
	+ \frac{1}{N} \left( W_t^n - \sum_{k=1}^N X_k - \int_0^t Q_s \, ds \right)^2     \,.
	\label{eq:approx}
\end{align}
Note that $Q_s \leq 2$. So if $T \ll \sqrt{N}$, 
the terms involving $Q_s$ in~\eqref{eq:approx} will vanish as $N\to\infty$.
Furthermore, if $X_k$ were I.I.D. random variables
with mean 0 and finite variance $\sigma^2_t$,
as $N\to \infty$, then the first two terms would converge to $0$ by the law of large numbers. This assumption is not strictly true, but by Theorem~\ref{theorem1} and~\eqref{eq:solutionV}, the correlation among all the random variables will vanish as $N\to\infty$ and for large time. Therefore we can take this as approximately true.

We consider the last term. By the law of the iterated logarithm~\cite{Feller1968, durrett2019probability}, we have that for large $N$
\begin{equation}
	\displaystyle\sum_{k=1}^N \frac{X_k }{\sqrt{2N}} \sim \sigma_t \sqrt{ \ln \ln N}.
	\label{eq:iteratedlog}
\end{equation}

Except for the largest $V^n$, which will stabilize to $2$, the rest of the $V^j$, where $j\not=n$, will go to zero.
By~\eqref{ine:correlation} and allowing for pathwise fluctuations,
in order to keep the product with the largest value on average to be~\eqref{ine:remark},
we expect that the trajectories that are attracted to 0 would
behave as
\begin{equation}
	V_j \sim \frac{1}{K_N + t } \,,
\end{equation}
where $K_N$ is some constant less than $(N-1)^2$ so that $1/K_N$ represents
the largest initial step size of $V_j$ over a unit of time at small $t$.

\begin{figure}[ht]
	\centering
	\includegraphics[width =0.8\linewidth]{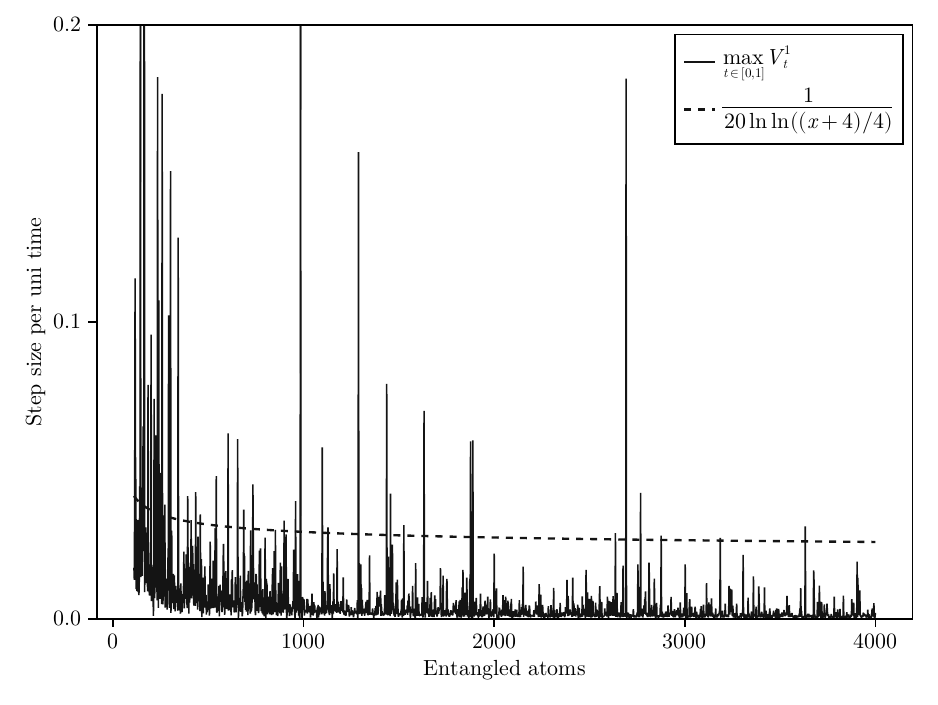}
	\caption{Path-wise initial step size for the first component $V_1$ with normal noise.
		The system of $N$ entangled atoms, $N =1, \dots, 4000$, is run until time $T = 1$.
		Vertical lines represent the running maximum of $V^1$ up until $T=1$.
		This tells us the step size over 1 unit of time of $V^1$.
		The dashed line is proportional to the inverse of the line that fits data of normal noise in Figure~\ref{fig-time}.}
	\label{fig-step}
\end{figure}

Let $T$ be the collapse time of the $N$ system. This is the time for
the maximum process $\displaystyle \max_{1\leq n \leq N} \left\{V_n\right\}$ to start from the initial position $2/N$ and cross  $2 - \delta$ for the first time.
The average step size over each unit of time of the maximum process $\max \left\{V_n\right\}_{n=1}^N$ is approximately $(2-\delta) / T$.
Initially almost all trajectories have equal step sizes per unit time.
So it may be reasonable to assume that
\begin{equation}
	\label{eq:initial}
	\frac{1}{K_N} \sim \frac{1}{T} \,,
\end{equation}
meaning
\begin{equation}
	K_N \sim T \,.
\end{equation}
This assumption is supported by Figure~\ref{fig-step}.

One can compute
\begin{align}
	\sigma_t^2 = \E \int_0^t (V^j_s)^2 \, ds & \sim \int_0^t \frac{1}{(K_N + s)^2 }  \, ds \nonumber \\
	                                         & \sim \frac{t}{T^2} \,.
	\label{ine:varestimate}
\end{align}
Now, as we would like to define collapse as when  $V^n \geq 2 - \delta$ for some $n = 1, \dots, N$, we can combine
\eqref{eq:approx}, \eqref{eq:iteratedlog}, and \eqref{ine:varestimate} to get
\begin{equation}
	2 - \delta \sim \sigma_T^2 \ln \ln N \sim  \frac{ 1}{T} \ln \ln N\,,
\end{equation}
where $T$ is the collapse time.
This leads to
\begin{equation}
	T \sim \ln \ln N \,,
\end{equation}
which is what we observe in the simulation.

We note that our argument is related but not the same as typical first passage time
argument~\cite{JordanKorotkov2010Uncollapsingwavefunctionundoing}.
The difference lies in the fact that we have $N$ correlated processes that are competing
to reach $2-\delta$ while typical first passage time argument applies to
the first time for a single process to cross certain threshold.
Even for $N$ independent processes, the analysis is vastly different and relies on
calculations related to large deviation principles~\cite{LawleyUniversalFormula2020} .
Our observation about the law of iterated logarithm in this work is surprising.

Lastly,  what presented here is far from a complete mathematical proof because
there are  certain points that that need to be carefully studied:
1) The I.I.D. nature of $V_n$'s that are not strictly true.
2) The bound~\eqref{ine:correlation} is not as sharp as it could be.
3) The initial largest step sizes over a unit of time of $V_n$'s is
a rough assumption.
If the $\ln \ln N$ bound is true and one hopes to prove it, we imagine that a generalized version of the law of the iterated logarithm for correlated random variables will
be needed in order to take care of the varying variance $\sigma_t$ as $N$ changes.
This is out of the scope of this present work.
Nevertheless, the heuristic argument presented here gives us
more confidence about the simulation result.

\bibliography{theBib, bibo}

\end{document}